# Universal Uncertainty Principle in Different Quantum Theories

C. Huang[1,2]    Yong-Chang Huang[3]

1. Lawrence Berkeley National Laboratory, 1 Cyclotron Road, Berkeley CA 94720, USA
2. Department of Physics and Astronomy, Purdue University, 525 Northwestern Avenue, W. Lafayette, IN 47907-2036, USA
3. Institute of Theoretical Physics, Beijing University of Technology, Beijing 100022, China

**Abstract**

This paper deduces universal uncertainty principle in different quantum theories after about one century of proposing uncertainty principle by Heisenberg, i.e., new universal uncertainty principle of any orders of physical quantities in quantum physics, overcomes the difficulty that current quantum computer, quantum communication, quantum control, quantum mechanics and so on theories cannot give exact values of general uncertainty of any orders of physical quantities, further gives all relevant different expressions of the universal uncertainty principle and their applications. In fact, our studies are consistent with current theories and physical factual experiments, e.g., relevant to hydrogen atom physics experiments. Using the new universal uncertainty principle, people can give all applications to atomic physics, quantum mechanics, quantum communication, quantum calculations, quantum computer and so on.

**Key words: quantum uncertainty principle, quantum calculation, quantum physics, quantum field, theoretical physics**

## I. Introduction

The uncertainty principle in quantum mechanics was first introduced in 1927 by W. Heisenberg, which shows that the more exactly the position of one particle is decided, the less exactly the momentum of the particle can be determined, and vice versa [1]. The current inequality depending on the standard deviation of position and momentum was deduced by E. H. Kennard later [2] and by H. Weyl in 1928 [3].

The uncertainty principle practically shows a basic physics law character of systems in quantum physics, and isn't one statement of the observational success of current technology [4].

Some pioneer researchers gave excellent investigations on uncertainty principle [5-9], Ref.[10] studied microscopic origin for the apparent uncertainty principle governing the anomalous attenuation, further, macroscopic quantum uncertainty principle and superfluid hydrodynamics were investigated [11], Ref.[12] researched on undecidability principle and the uncertainty principle even for classical systems. Ref.[13] studied hierarchy of local minimum solutions of Heisenberg's uncertainty principle, furthermore, noncommutative spacetime, stringy spacetime uncertainty principle, and density fluctuations were shown [14].

A. Bina, S. Jalalzadeh and A. Moslehi showed a quantum black hole in the generalized uncertainty principle framework [15], and Ref.[16] gave a unification theory of classical statistical uncertainty relation and quantum uncertainty relation and its applications. Pierre Nataf, Mehmet Dogan and Karyn Le Hur further revealed Heisenberg uncertainty principle as a probe of entanglement entropy: Application to superradiant quantum phase transitions [17].

Giuseppe Vallone, Davide G. Marangon, Marco Tomasin and Paolo Villoresi show quantum randomness certified by the uncertainty principle [18], Smail Bougouffa and Zbigniew Ficek further reveal evidence of indistinguishability and entanglement determined by the energy-time uncertainty principle in a system of two strongly coupled bosonic modes [19]. Furthermore, L. Perivolaropoulos gives research on cosmological horizons, uncertainty principle, and maximum length quantum mechanics [20].

Shmuel Friedland, Vlad Gheorghiu, and Gilad Gour give universal uncertainty relations [21], and Andre C. Barato and Udo Seifert further present thermodynamic uncertainty relation for biomolecular processes [22].

Since we study the general physical laws of any quantum systems of arbitrary space, so far there is no the corresponding arbitrary uncertainty principle, so we need to study and give arbitrary uncertainty principle.

So far in quantum physics, all the uncertainty relations are, for the highest considering, the uncertainty relationship about variance level, higher order uncertainty relation cannot be given out, which leads to, when doing the measurement of any orders of two arbitrary physical quantities in the real cases, various statistical accuracies are not high, the most general estimations and calculations of physics quantities cannot be done for general different physics systems.

Because this paper is concerned with any physical system of arbitrary space, so we have to generalize the original uncertain principle to the most general case, therefore, in this paper, we will solve the theory and representation problem of new any order uncertainty principle of any two physical quantities, which cannot be solved so far.

This paper is arranged as follows: Sect. two gives a new general unequal theorem in mathematics and a general useful inequality of any orders of functions of two operators; Sect. three is new universal uncertainty principle of general physics quantities in quantum physics; Sect. four shows research on a moving particle with mass in a general central force field; Sect. five presents more applications of this theory; Sect. six shows new general uncertainty relations of general physics quantities in classic statistics; Sect. seven gives summary, conclusion and outlook.

## II. A new general unequal theorem in mathematics and a general useful inequality of any orders of functions of two operators

For a $\mu$-integrable function $f$ and $p > 0$, then $|f|^p$ is $\mu$-integrable [23].

Evidently, if $f$ is $\mu$-integrable and $0 \leq \rho(\mu) = \psi^+(\mu)\psi(\mu) \leq 1$, one has

$$\left| \int_\Omega f(\mu)\rho(\mu)d\mu \right|^p = \left| \int_\Omega \psi^+(\mu)f(\mu)\psi(\mu)d\mu \right|^p \leq \int_\Omega \psi^+(\mu)|f(\mu)|^p \psi(\mu)d\mu . \quad (2.1)$$

Defining

$$\|f(\mu)\|_p = (\int_\Omega \psi^+(\mu)|f(\mu)|^p \psi(\mu)d\mu)^{1/p}, \quad (2.2)$$

where $\int_\Omega \psi^+(\mu)|f(\mu)|^p \psi(\mu)d\mu < \infty$, in order to study the new general essential theorem, we first use the following lemma [24]:

**Lemma**. If $0 < c, d, p, q < \infty$ and $r = \dfrac{1}{p} + \dfrac{1}{q} = \dfrac{p+q}{pq}$, there must be

$$\frac{c^p}{p} + \frac{d^q}{q} \geq r c^{\frac{1}{r}} d^{\frac{1}{r}}. \tag{2.3}$$

Using inequality (2.3), we can show a new general essential functional unequal theorem.

**Unequal theorem**. Let $0 < p, q < \infty$ and $r = \dfrac{1}{p} + \dfrac{1}{q}$, If $f$ and $g$ are $\mu$-integrable functions, there must exist

$$\int_\Omega \psi^+ |fg|^{\frac{qp}{q+p}} \psi d\mu \leq (\int_\Omega \psi^+ |f|^p \psi d\mu)^{\frac{q}{q+p}} (\int_\Omega \psi^+ |g|^q \psi d\mu)^{\frac{p}{q+p}}. \tag{2.4}$$

Proof. In the Lemma, taking $c = \dfrac{|f(\mu)|}{\|f(\mu)\|_p}, d = \dfrac{|g(\mu)|}{\|g(\mu)\|_q}$ and substituting them into inequality (2.3), we have

$$r \frac{|f(\mu)g(\mu)|^{\frac{1}{r}}}{\|f(\mu)\|_p^{\frac{1}{r}} \|g(\mu)\|_q^{\frac{1}{r}}} \leq \frac{|f(\mu)|^p}{p\|f(\mu)\|_p^p} + \frac{|g(\mu)|^q}{q\|g(\mu)\|_q^q}. \tag{2.5}$$

Multiplying inequality (2.5) with $\psi^+(\mu)$ and $\psi(\mu)$ from the left and right sides, respectively and integrating, thus we obtain inequality (2.4).

For a general relation between quantum physics and classic physics, physical consistent property demands that there is a fundamental result that the quantum average value of any operator in quantum physics is equal to the relative statistical average value of the classic physics quantity corresponding to its quantum operator, which can be mathematically expressed as [16,25]

$$\int_\Omega X\rho d\mu = \int_\Omega \psi^+ X\psi d\mu = \int_\Omega \psi^+ \hat{X}\psi d\mu, \quad \rho = \psi^+\psi, \tag{2.6}$$

where $\hat{X}$'s concrete function expression (e.g., $\hat{\vec{p}} = -i\hbar\nabla$) is generally different from $X$'s concrete function expression (e.g., $\vec{p} = m\dot{\vec{r}}$), but their average values are the same, which is just physics consistence demand; for any classic real density, it always can be written as the product of a complex function and its complex conjugate function, i.e., $\rho = \psi^+\psi$, and we take the complex function as the corresponding wave function of quantum mechanics. Vice versa, i.e., we can take the product of a quantum wave function and its complex conjugate wave function as classic density, namely, $\rho = \psi^+\psi$.

When statistical density $\rho$ of classic physics is not dependent on the corresponding classic physics quantity, then the statistical average value of the corresponding quantum physics quantity is just degenerated from Eq.(2.6) as the value of the corresponding classic physics quantity [16,25]. i.e.,

$$X = \bar{\hat{X}} = \int_\Omega \psi^+ \hat{X} \psi d\mu, \int_\Omega \rho d\mu = 1. \tag{2.7}$$

Because $\hat{X}$ is any operator, which may naturally be a composite function of operators $\hat{A}, \hat{B}$, namely, $\hat{X} = \hat{X}(\hat{A}, \hat{B})$, then Eq.(2.6) can be rewritten as

$$\int_\Omega \psi^+ X(A,B) \psi d\mu = \int_\Omega \psi^+ \hat{X}(\hat{A}, \hat{B}) \psi d\mu. \tag{2.8}$$

Eq.(2.8) just is of the basic demand of physical consistent property.

For any positive real number $k$, when especially taking $\hat{X}(\hat{A}, \hat{B}) = \left|\hat{f}(\hat{A}, \hat{B})\right|^k$, Eq.(2.8) is degenerated as

$$\int_\Omega \psi^+ |f(A,B)|^k \psi d\mu = \int_\Omega \psi^+ \left|\hat{f}(\hat{A}, \hat{B})\right|^k \psi d\mu. \tag{2.9}$$

where $f$ is a general function of classic physics quantities $A$ & $B$, concrete function expressions of $\hat{f}(\hat{A}, \hat{B})$ & $f(A,B)$ are generally different with each other as usual, e.g., $\hat{f}(\hat{A}, \hat{B}) = \frac{\hat{A}\hat{B} + \hat{B}\hat{A}}{2} \frac{\hat{B}}{2} + \frac{\hat{B}}{2} \frac{\hat{A}\hat{B} + \hat{B}\hat{A}}{2} = \frac{\hat{A}\hat{B}^2 + 2\hat{B}\hat{A}\hat{B} + \hat{B}^2\hat{A}}{4}$ is different from $f(A,B) = AB^2$; and the absolute value mark in right hand side of Eq.(2.9) means that after the operators have acted on the wave function, one should concretely take the whole obtained values contributed from the operators as modulus value, due to $\left|\hat{f}(\hat{A}, \hat{B})\right|^k$ operators' action chacracter necesarily acting on wave function so that the average value expression of the opreator function in Eq.(2.9) is equal to the average value expression of its classic physics quantity function, which is just the physics consistence demand and can give a constrainting equation determinting the correspondenc between the quantum system and its classic physics system. For exmple, a concrete application can be found in inequality (5.2).

Using inequality (2.4) and Eq.(2.9), we can get a general useful inequality

$$\int_\Omega \psi^+ \left|\hat{f}\hat{g}\right|^{\frac{qp}{q+p}} \psi d\mu \leq (\int_\Omega \psi^+ \left|\hat{f}\right|^p \psi d\mu)^{\frac{q}{q+p}} (\int_\Omega \psi^+ \left|\hat{g}\right|^q \psi d\mu)^{\frac{p}{q+p}}. \tag{2.10}$$

From the above research, it can be seen that inequality (2.10) is general, exact and perfect in both mathematical deduction and physics research. Using inequality (2.10), we can do a lot of investigations in different quantum theories.

## III. New universal uncertainty principle of general physics quantities in quantum physics

From current quantum uncertainty principle mathematics derivation and analysis of the entire physics' building process, only simply taking Schwarz inequality related to the variance of two physical quantities in the uncertainty description of the same time measurement is just a special case, because, according to a general principle, a real system of quantum physics also abounds non-Schwarz inequality to calculate measurement uncertainty description of the variance of two physical quantities at the same time, the variance description relating to Schwarz inequality is just a special case of them. Consequently, the current quantum theory about the description of the uncertainty principle is not comprehensive, not perfect and not strict enough, to overcome these shortcomings, we now need to put it to the most general case

Now we use inequality (2.10) to deduce new uncertainty principle of any orders of two physical quantities in quantum physics in following studies.

For any operators $\hat{A}, \hat{B}$, we take $\hat{f} = \Delta\hat{A} = \hat{A}-<\hat{A}>, \hat{g} = \Delta\hat{B} = \hat{B}-<\hat{B}>$ and substituting them into inequality (2.10), we have

$$\int_\Omega \psi^+ \left|\Delta\hat{A}\Delta\hat{B}\right|^{\frac{qp}{q+p}} \psi d\mu = \frac{1}{2^{\frac{qp}{q+p}}} \int_\Omega \psi^+ \left|[\Delta\hat{A},\Delta\hat{B}]+\{\Delta\hat{A},\Delta\hat{B}\}\right|^{\frac{qp}{q+p}} \psi d\mu$$
$$\leq (\int_\Omega \psi^+ \left|\Delta\hat{A}\right|^p \psi d\mu)^{\frac{q}{q+p}} (\int_\Omega \psi^+ \left|\Delta\hat{B}\right|^q \psi d\mu)^{\frac{p}{q+p}} \quad . \quad (3.1)$$

For Hermite operator $\{\Delta\hat{A}, \Delta\hat{B}\} = \hat{X}_H$ and anti-Hermite operator $[\Delta\hat{A}, \Delta\hat{B}] = [\hat{A}, \hat{B}] = \hat{Y}_{aH}$, we generally consider that there are their eigenstates $|\Phi_X\rangle$ and $|\Phi_Y\rangle$ such that $\hat{X}_H|\Phi_X\rangle + \hat{Y}_{aH}|\Phi_Y\rangle = X_H|\Phi_X\rangle + iY_{aH}|\Phi_Y\rangle$ (because anti-Hermite operator eigenvalue is imaginary number), where $X_H$ and $Y_{aH}$ are real values, using general complete normalization conditions $\sum_x |\Phi_X(x)\rangle\langle\Phi_X(x)| = 1 (= \int |\Phi_X(x)\rangle\langle\Phi_X(x)|dx$ for continuous variables) and $\sum_y |\Phi_Y(y)\rangle\langle\Phi_Y(y)| = 1 (= \int |\Phi_Y(y)\rangle\langle\Phi_Y(y)|dy$ for continuous variables), namely, using the path integral quantization method [26], then inequality (3.1) can be again expressed as

$$\frac{1}{2^{\frac{qp}{q+p}}} \int_\Omega \psi^+ \left|\{\Delta\hat{A},\Delta\hat{B}\}+[\hat{A},\hat{B}]\right|^{\frac{qp}{q+p}} \psi d\mu$$

$$= \frac{1}{2^{\frac{qp}{q+p}}} \int_\Omega \psi^+ \left|\int \hat{X}_H |\Phi_X(x)\rangle\langle\Phi_X(x)| dx + \int \hat{Y}_{aH} |\Phi_Y(y)\rangle\langle\Phi_Y(y)| dy\right|^{\frac{qp}{q+p}} \psi d\mu \quad (3.2)$$

$$= \frac{1}{2^{\frac{qp}{q+p}}} \int_\Omega \psi^+ \left|\int X_H(x) |\Phi_X(x)\rangle\langle\Phi_X(x)| dx + i\int Y_{aH}(y) |\Phi_Y(y)\rangle\langle\Phi_Y(y)| dy\right|^{\frac{qp}{q+p}} \psi d\mu$$

$$\leq (\int_\Omega \psi^+ |\Delta\hat{A}|^p \psi d\mu)^{\frac{q}{q+p}} (\int_\Omega \psi^+ |\Delta\hat{B}|^q \psi d\mu)^{\frac{p}{q+p}}$$

For a general complex $c=a+ib$, one has $|c|=|a+ib|\geq|b|$, thus using inequality (3.2), we can similarly have

$$\frac{1}{2^{\frac{qp}{q+p}}} \int_\Omega \psi^+ \left|\int Y_{aH}(y) |\Phi_Y(y)\rangle\langle\Phi_Y(y)| dy\right|^{\frac{qp}{q+p}} \psi d\mu$$

$$= \frac{1}{2^{\frac{qp}{q+p}}} \int_\Omega \psi^+ \left|\hat{Y}_{aH} \int |\Phi_Y(y)\rangle\langle\Phi_Y(y)| dy\right|^{\frac{qp}{q+p}} \psi d\mu \quad . \quad (3.3)$$

$$\leq (\int_\Omega \psi^+ |\Delta\hat{A}|^p \psi d\mu)^{\frac{q}{q+p}} (\int_\Omega \psi^+ |\Delta\hat{B}|^q \psi d\mu)^{\frac{p}{q+p}}$$

Therefore, we achieve

$$\frac{1}{2^{\frac{qp}{q+p}}} \int_\Omega \psi^+ \left|[\hat{A},\hat{B}]\right|^{\frac{qp}{q+p}} \psi d\mu \leq (\int_\Omega \psi^+ |\Delta\hat{A}|^p \psi d\mu)^{\frac{q}{q+p}} (\int_\Omega \psi^+ |\Delta\hat{B}|^q \psi d\mu)^{\frac{p}{q+p}}. \quad (3.4)$$

where inequality (3.4) is stronger inequality than inequality (3.2), because we have deleted the first term in the third line in inequality (3.2).

Inequality (3.4) is just the new universal uncertainty principle of any orders of two physical quantities.

When $p=q$, inequality (3.4) can be further simplified as

$$\frac{1}{2^{\frac{p}{2}}} \int_\Omega \psi^+ \left|[\hat{A},\hat{B}]\right|^{\frac{p}{2}} \psi d\mu \leq (\int_\Omega \psi^+ |\Delta\hat{A}|^p \psi d\mu)^{\frac{1}{2}} (\int_\Omega \psi^+ |\Delta\hat{B}|^p \psi d\mu)^{\frac{1}{2}}. \quad (3.5)$$

When $p=2$, inequality (3.5) can be again simplified as

$$\frac{1}{2} \int_\Omega \psi^+ \left|[\hat{A},\hat{B}]\right| \psi d\mu \leq (\int_\Omega \psi^+ |\Delta\hat{A}|^2 \psi d\mu)^{\frac{1}{2}} (\int_\Omega \psi^+ |\Delta\hat{B}|^2 \psi d\mu)^{\frac{1}{2}}. \quad (3.6)$$

Inequality (3.6) is just the old well-known usual uncertainty principle. Therefore, our studies are consistent with current theory.

**IV. Research on a generally moving particle with a general mass in general central force field in quantum mechanics**

For a generally moving particle with a general mass in general central force field in quantum mechanics

$$V = -\frac{\beta}{r^\alpha}, \beta > 0, \quad (4.1)$$

using Virial theorem, we have the mean value relationship between kinetic energy and potential energy in any bound state

$$<T> = -\frac{1}{2}\overline{\vec{F}\cdot\vec{r}} = \frac{1}{2}<\vec{r}\cdot\nabla V> = -\frac{1}{2}<r\frac{d}{dr}\frac{\beta}{r^\alpha}> = -\frac{\alpha}{2}<V>. \quad (4.2)$$

Thus we have total energy

$$E = <T+V> = (1-\frac{2}{\alpha})<T> = (1-\frac{\alpha}{2})<V> = (\frac{\alpha}{2}-1)\beta<\frac{1}{r^\alpha}>, \quad (4.3)$$

when $0 < \alpha < 2$, Eq.(4.3) take a negative value, then the system has the different bounded states, i.e., this system may form stable matter state.

On the other hand, using the deduced inequality (2.10) and taking $f = r, g = r^{-1}$, we deduce

$$\int_\Omega \psi^+ \left|rr^{-1}\right|^{\frac{qp}{q+p}} \psi d\mu = 1 \leq (\int_\Omega \psi^+ |r|^p \psi d\mu)^{\frac{q}{q+p}} (\int_\Omega \psi^+ |r^{-1}|^q \psi d\mu)^{\frac{p}{q+p}} \quad . \quad . \quad (4.4)$$

Thus we generally deduce

$$(\int_\Omega \psi^+ |r^{-1}|^q \psi d\mu)^{\frac{p}{q+p}} \geq \frac{1}{(\int_\Omega \psi^+ |r|^p \psi d\mu)^{\frac{q}{q+p}}}, \quad . \quad (4.5)$$

inequality (4.5) is new inequality that cannot be deduced in the past, gives new physics and is very useful and important in quantum theory, quantum field theory, atomic physics etc., i.e., inequality (4.5) gives the new physics connection between $(\int_\Omega \psi^+ |r^{-1}|^q \psi d\mu)^{\frac{p}{q+p}}$ and $\frac{1}{(\int_\Omega \psi^+ |r|^p \psi d\mu)^{\frac{q}{q+p}}}$ for a generally moving particle with a general mass in general central force field, which cannot be given in current quantum mechanics. When $p=q$, inequality (4.5) is further simplified as

$$(\int_\Omega \psi^+ |r^{-1}|^p \psi d\mu)^{\frac{1}{2}} \geq \frac{1}{(\int_\Omega \psi^+ |r|^p \psi d\mu)^{\frac{1}{2}}}, \quad . \quad (4.6)$$

when $p = \alpha$ and $r$ & $r^{-1}$ are bigger than zero, inequality (4.6) is finally simplified as

$$<\frac{1}{r^\alpha}>^{1/2} \geq \frac{1}{<r^\alpha>^{1/2}} \quad . \quad (4.7)$$

Inequality (4.5) is, for the first time, achieved, which can be extensively applied to a lot of problems of central force fields in quantum mechanics, atomic physics and so on.

Using Eqs.(4.1) and inequality (4.7), we deduce

$$E = <T+V> = <\frac{\hat{\vec{P}}^2}{2m}> + <-\frac{\beta}{r^\alpha}> \leq <\frac{\hat{\vec{P}}^2}{2m}> - \frac{\beta}{<r^\alpha>}. \qquad (4.8)$$

On the other hand, using uncertainty relation

$$\Delta r \Delta p_r \geq \frac{\hbar}{2}, \qquad (4.9)$$

and for the bound states in quantum mechanics, one has

$$<\vec{p}> = 0, \Delta \vec{p}^2 = \vec{p}^2, \quad p_r^2 = \Delta p_r^2 \geq \frac{\hbar^2}{4\Delta r^2}. \qquad (4.10)$$

Using Eq.(4.10), for ground state, we can only consider $p_r$ component, then expression (4.8) can be further expressed as

$$\begin{aligned} E = <T+V> &= \frac{<p_r^2>}{2m} + <V> \\ &\geq \frac{\hbar^2}{8m} \frac{1}{<r^2> - <r>^2} - \beta <\frac{1}{r^\alpha}> \\ &\leq \frac{\hbar^2}{8m} \frac{1}{<r^2> - <r>^2} - \beta \frac{1}{<r^\alpha>} \end{aligned} \qquad (4.11)$$

where the last line uses inequality (4,7).

For the bound state, expression (4.11) also needs to be smaller than or equate zero. Thus taking inequality (4.11) equal to zero, we have

$$\frac{\hbar^2}{8m} \frac{1}{<r^2> - <r>^2} = \beta \frac{1}{<r^\alpha>}, or <r^\alpha> = \frac{8m\beta}{\hbar^2}(<r^2> - <r>^2), \qquad (4.12)$$

where $0 < \alpha < 2$, the different $\alpha$ values give different central force field characters, for $\alpha = 1$, which is just the case we are very familiar with [25], and $\beta$ may be, e.g., $\beta = K_{Q_1Q_2}|Q_1Q_2|$, $Q_1, Q_2$ are any charges the new any system is with, e.g., some kinds of electric charges, magnetic charges, mass charges and so on [27], $K_{Q_1Q_2}$ is the corresponding coupling coefficient.

When $\alpha = 1$, Eq.(4.12) can be rewritten as

$$\begin{aligned} b<r>^2 + <r> - b<r^2> &= 0 \\ <r> &= \frac{-1}{2b} \pm \sqrt{1/(4b^2) + <r^2>} \end{aligned}, \qquad (4.13)$$

where $b = \frac{8m\beta}{\hbar^2}$, Eq.(4.13) shows that $<r>$ is a function of $<r^2>$ and relevant parameters $\frac{8m\beta}{\hbar^2}$ when the bound state is disintegrated. Thus, expressions (4.11)-(4.13) have extensive uses in quantum mechanics, atomic physics and so on, because they are very general expressions, which cannot be given in the current quantum theories.

For example, we consider the most common expression of Lennard-Jones potential ( a useful model for the interaction of a pair of neutral atoms or molecules ) [28]

$$V_{LJ} = 4\varepsilon[(\frac{\sigma}{r})^{12} - (\frac{\sigma}{r})^{6}],  \qquad (4.14)$$

where $\sigma$ is a distance of the inter-particle potential being zero, $\varepsilon$ is a depth of a potential well, $r$ is a distance between the two particles, the Lennard-Jones potential is extensively and accurately used in computer simulations [28].

The $r^{-12}$ term describes, at short ranges, Pauli repulsion due to overlapping electron orbitals, and the $r^{-6}$ term describes, at long ranges, attraction (dispersion force or van der Waals force).

Potential (4.14) was improved later as [29]

$$V_B = \gamma[e^{-r/r_0} - (\frac{\sigma}{r})^{6}], \qquad (4.15)$$

it is a very good and useful potential at long and short distances for neutral atoms and molecules.

Using Eq.(4.15), we have

$$<V_B> = \gamma[<e^{-r/r_0}> - <(\frac{\sigma}{r})^{6}>]$$
$$\leq \gamma[<e^{-r/r_0}> - \frac{\sigma^6}{<r^6>}], \qquad (4.16)$$

where we used expression (4.7). Expression (4.16) gives a quantum average max value for the improved potential, which cannot be obtained in the past quantum theories.

## V. More applications of this theory

Using Eq.(2.9), we have

$$\frac{1}{2^{\frac{qp}{q+p}}} \int_\Omega \psi^+ \left|[\hat{A},\hat{B}]\right|^{\frac{qp}{q+p}} \psi d\mu = \frac{1}{2^{\frac{qp}{q+p}}} \int_\Omega \psi^+ \left|\{A,B\}i\hbar\right|^{\frac{qp}{q+p}} \psi d\mu$$
$$= (\frac{\hbar}{2})^{\frac{qp}{q+p}} \int_\Omega \psi^+ \left|\{A,B\}\right|^{\frac{qp}{q+p}} \psi d\mu \qquad (5.1)$$

where $\{A,B\}(=\frac{[\hat{A},\hat{B}]}{i\hbar})$ is Poisson bracket in classic analytical mechanics, which means when quantum quantities return to classic quantities, people need to change the quantum commutative relation to the corresponding classic commutative relation. These changes guarantee that our investigations are consistent with all current relevant theories.

When $\hat{A} = x_i, \hat{B} = \hat{p}_j$, inequality (3.4) can be simplified as

$$\frac{1}{2^{\frac{qp}{q+p}}} \int_\Omega \psi^+ \left|[\hat{x}_i, \hat{p}_j]\right|^{\frac{qp}{q+p}} \psi d\mu = \frac{1}{2^{\frac{qp}{q+p}}} \int_\Omega \psi^+ \left|i\hbar \delta_{ij}\right|^{\frac{qp}{q+p}} \psi d\mu$$
$$= (\frac{\hbar}{2})^{\frac{qp}{q+p}} \delta_{ij} \leq (\int_\Omega \psi^+ |\Delta \hat{x}_i|^p \psi d\mu)^{\frac{q}{q+p}} (\int_\Omega \psi^+ |\Delta \hat{p}_j|^q \psi d\mu)^{\frac{p}{q+p}}. \quad (5.2)$$

When $p=q$, inequality (5.2) can be further simplified as

$$(\frac{\hbar}{2})^{\frac{p}{2}} \delta_{ij} \leq (\int_\Omega \psi^+ |\Delta \hat{x}_i|^p \psi d\mu)^{\frac{1}{2}} (\int_\Omega \psi^+ |\Delta \hat{p}_j|^p \psi d\mu)^{\frac{1}{2}}. \quad (5.3)$$

When $p=2$, inequality (5.3) can be further simplified as

$$\frac{\hbar}{2} \delta_{ij} \leq (\int_\Omega \psi^+ |\Delta \hat{x}_i|^2 \psi d\mu)^{\frac{1}{2}} (\int_\Omega \psi^+ |\Delta \hat{p}_j|^2 \psi d\mu)^{\frac{1}{2}}. \quad (5.4)$$

Inequality (5.4) is just the current well-known uncertainty principle between $x_i$ and $\hat{p}_j$. Therefore, our studies are again concretely consistent with current theory.

For example, we further consider the hydrogen atom basic state $\psi_{100} = (\frac{1}{\pi a_0^3})^{1/2} e^{-r/a_0}$, due to the sphere symmetry property of the basic state $\psi_{100}$, we easily obtains $\overline{\hat{x}_i} = \overline{\hat{p}_i} = 0$ ($i=1,2,3$), substituting these zero average values into inequality (5.2), we have

$$(\int_\Omega \psi^+ |\hat{x}_i|^p \psi d\mu)^{\frac{q}{q+p}} (\int_\Omega \psi^+ |\hat{p}_j|^q \psi d\mu)^{\frac{p}{q+p}} \geq (\frac{\hbar}{2})^{\frac{qp}{q+p}} \delta_{ij}. \quad (5.5)$$

No losing generality, taking $p=3, q=2$ in inequality (5.5), we get

$$(\frac{1}{\pi a_0^3})^{2/5} (\int_\Omega |\hat{x}_i|^3 e^{-2r/a_0} d\mu)^{\frac{2}{5}} (\int_\Omega \psi^+ |\hat{p}_j|^2 \psi d\mu)^{\frac{3}{5}} \geq (\frac{\hbar}{2})^{\frac{6}{5}} \delta_{ij}. \quad (5.6)$$

For simplicity and no losing generality, taking $i=j=3, k = 2/a_0$ and calculating the first integration, we have

$$\int_\Omega |\hat{x}_3|^3 e^{-2r/a_0} d\mu = -\int_\Omega |r\cos\theta|^3 e^{-2r/a_0} r^2 d\cos\theta dr d\phi$$
$$= -(\int_0^{\pi/2} + \int_{\pi/2}^{\pi}) \int_0^\infty |\cos\theta|^3 e^{-kr} r^5 dr d\cos\theta 2\pi$$
$$= -(\int_0^{\pi/2} d\cos^4\theta/4 - \int_{\pi/2}^{\pi} d\cos^4\theta/4) \int_0^\infty e^{-kr} r^5 dr 2\pi$$
$$= -(-1/4 - 1/4) \int_0^\infty e^{-kr} r^5 dr 2\pi = \pi \int_0^\infty e^{-kr} r^5 dr$$

$$= \pi(-\frac{d}{dk})^5 \int_0^\infty e^{-kr} dr = -\pi(\frac{d}{dk})^5 \frac{1}{k}$$
$$= -\pi(-1)(-2)(-3)(-4)(-5)k^{-6} = \frac{15\pi}{8} a_0^6 \tag{5.7}$$

Calculating the second integration, we obtain

$$\int_\Omega \psi^+ |\hat{p}_j|^2 \psi d\mu = \frac{1}{3}\int_\Omega \psi^+ |\vec{\hat{p}}|^2 \psi d\mu = \hbar^2 \frac{1}{3}\int_\Omega |\nabla \psi_{100}|^2 d\mu = \frac{\hbar^2}{3a_0^2}, \tag{5.8}$$

where $j = 1,2,3$, and we have used the sphere symmetry property of the basic state $\psi_{100}$.

Putting expressions (5.7) and (5.8) into expression (5.6), we can deduce

$$(\frac{1}{\pi a_0^3})^{2/5} (\int_\Omega |z|^3 e^{-2r/a_0} d\mu)^{\frac{2}{5}} (\int_\Omega \psi^+ |\hat{p}_z|^2 \psi d\mu)^{\frac{3}{5}}$$
$$= (\frac{1}{\pi a_0^3})^{2/5} (\frac{15\pi}{8} a_0^6)^{\frac{2}{5}} (\frac{\hbar^2}{3a_0^2})^{\frac{3}{5}} \geq (\frac{\hbar}{2})^{\frac{6}{5}} \tag{5.9}$$

Inequality (5.9) can be simplified as

$$\sqrt[5]{15^2 (\frac{1}{3})^3} = \sqrt[5]{\frac{25}{3}} > 1 \tag{5.10}$$

Thus we see that not only all quantities are with the same dimensions between two sides of inequality (5.9), but also their coefficients of two sides satisfy the inequality.

Consequently, it can be seen that inequality (5.10) just shows the correction of inequality (5.6) in real physical experiments.

These are consistent with all the known results and include the achieved successes in quantum theory.

Using inequality (3.4), we can give all uncertainty relations of any orders of two physical quantities, which can be applied to atomic physics, quantum mechanics, quantum communications, quantum calculations, quantum computer and so on, e.g., to Refs. [12-21].

## VI. New general uncertainty relations of general physics quantities in classic statistical physics

Using the unequal theorem, when $f$ and $g$ are classic physics quantities, and taking $\rho = \psi^+ \psi$ density function, then unequal expression (2.4) can be rewritten as

$$\int_\Omega |fg|^{\frac{qp}{q+p}} \rho d\mu \leq (\int_\Omega |f|^p \rho d\mu)^{\frac{q}{q+p}} (\int_\Omega |g|^q \rho d\mu)^{\frac{p}{q+p}}. \tag{6.1}$$

When $p=q$, inequality (6.1) can be further simplified as

$$\int_\Omega |fg|^{\frac{p}{2}} \rho d\mu \leq (\int_\Omega |f|^p \rho d\mu)^{\frac{p}{2}} (\int_\Omega |g|^p \rho d\mu)^{\frac{p}{2}}. \tag{6.2}$$

When $p=2$, inequality (6.2) can be again simplified as

$$\int_\Omega |fg|\rho d\mu \leq \int_\Omega |f|^2 \rho d\mu \int_\Omega |g|^2 \rho d\mu, \qquad . \qquad (6.3)$$

Further when taking $\rho = 1$, inequality (6.3) is reduced as

$$\int_\Omega |fg| d\mu \leq \int_\Omega |f|^2 d\mu \int_\Omega |g|^2 d\mu. \qquad . \qquad (6.4)$$

Inequality (6.4) is just the well-known Schwarz inequality. Therefore, our studies are consistent with current statistical theory.

## VII. Summary, conclusion and outlook

We first simply review the development of uncertainty principle, and then use a general unequal lemma to deduce a general unequal theorem of any orders of any two functions with integrations.

This paper deduces an important relation between a composite function of operators in quantum physics and a composite function of classic physics quantities corresponding the operators, further utilizes the relation and the unequal theorem to achieve a general important inequality. Using the inequality, we show new universal uncertainty principle in different quantum theories, namely, new universal uncertainty principle of any orders of physical quantities in quantum physics, overcome the difficulty that current quantum computer, quantum communication, quantum control and so on theories cannot give exact values of general uncertainty of any orders of physical quantities, further, we give all relevant different expressions of the universal uncertainty principle and their applications.

For examples, inequality (3.4) is just the new uncertainty principle of any orders of physical quantities in quantum physics. When $p=q$, inequality (3.4) can be further simplified as inequality (3.5); when $p=2$, inequality (3.5) can be further simplified as the well-known usual uncertainty principle expression. Therefore, our studies are consistent with current theory and real physical experiments, e.g., relevant to hydrogen atom physics experiments.

For a moving particle with mass in a general central force field, using Virial theorem, we achieve a new general average value relationship (4.2) between kinetic energy and potential energy in any bound state, thus we deduce the total energy (4.3) proportional to $<1/r^\alpha>$ ($0<\alpha<2$).

Furthermore, for the general potential, we generally deduce a new radial general interesting inequality $<r^{-q}>^{\frac{p}{q+p}} \geq 1/<r^p>^{\frac{q}{q+p}}$ (4.5), which cannot be deduced in the past and can give many new applications in relevant redial calculations in different quantum theories.

We further achieve a new key important inequality $<|\Delta\hat{x}_i|^p>^{\frac{q}{q+p}} <|\Delta\hat{p}_j|^q>^{\frac{p}{q+p}} \geq \delta_{ij}(\frac{\hbar}{2})^{\frac{qp}{q+p}}$ ($0 < $ p, q $ < \infty$), when $p=q=2$, which is just the well-known usual uncertainty relation between $x_i$ and $\hat{p}_j$. Therefore, our studies are again concretely consistent with current theory.

On the other hand, using the deduced unequal theorem, unequal expression (2.4) can be simplified as inequality (6.1), when *p=q=2*, inequality (6.1) can be further simplified as the well-known Schwarz inequality. Therefore, our studies are consistent with current functional theory.

Using the new universal uncertainty principle expression (3.4), we can give all uncertainty relations of any orders of physical quantities in quantum physics, which can be applied to quantum mechanics, quantum communication, quantum calculations and so on, because, in these theories, they all use quantum theories and the uncertainty relation to investigate and solve their relevant problems.

Therefore, this paper opens a new area investigating new universal uncertainty principle of any orders of physics quantities in quantum physics, the achieved results in this paper are useful, because this paper will influence the research of others in physics.

For any statistical system, it is usually necessary to give a description of all high order statistical moments in order to give a rigorous statistical description theory [30]. The same is true for any quantum statistical physics system [25,30]. In the current theory of quantum statistical physics systems, there is only the same time measurement uncertainty description of the variance of two physical quantities [25, 30], and a lot of the rests of the more, stricter and more general descriptions so far has not been given. This paper just gives these descriptions that so far have still not been given, which overcome the difficulty that it is not possible to strictly describe the physical systems related to the aspect of the universal uncertainty principle, so as to give a new general description theory of the quantum physics relating to the universal uncertainty principle.

**Acknowledgments:** The work is supported by the U.S. Department of Energy, contract no. DE-AC02-05CH11231, NSF through grants PHY-08059, DOE through grant DEFG02-91ER40681 and National Natural Science Foundation of China (No. 11173028 and No.10875009).

# References

[1] W. Heisenberg, "Über den anschaulichen Inhalt der quantentheoretischen Kinematik und Mechanik", Zeitschrift für Physik (in German), 43 (3–4) (1927) 172–198, Bibcode:1927ZPhy...43..172H, doi:10.1007/BF01397280. Annotated pre-publication proof sheet of Über den anschaulichen Inhalt der quantentheoretischen Kinematik und Mechanik, March 21, 1927.

[2] Kennard, E. H. (1927), "Zur Quantenmechanik einfacher Bewegungstypen", Zeitschrift für Physik (in German), 44 (4–5): 326–352, Bibcode:1927ZPhy...44..326K, doi:10.1007/ BF01391200.

[3] H. Weyl, Gruppentheorie und Quantenmechanik, Leipzig: Hirzel (1928).

[4] Indian Institute of Technology Madras, Professor V. Balakrishnan, Lecture 1 – Introduction to Quantum Physics; Heisenberg's uncertainty principle, National Programme of Technology Enhanced Learning on YouTube; L. D. Landau, E. M. Lifshitz, Quantum Mechanics: Non-Relativistic Theory. Vol. 3 (3rd ed.). Pergamon Press (1977). ISBN 978-0-08-020940-1.

[5] G. Breit, The Principle of Uncertainty in Weyl's System, Phys. Rev. 32, 570 (1928)

[6] H. P. Robertson, The Uncertainty Principle, Phys. Rev. 34, 163 (1929).

[7] Alexander W. Stern, The Uncertainty Principle, Phys. Rev. 37, 1186 (1931).

[8] N. Rosen and M. S. Vallarta, Relativity and the Uncertainty Principle, Phys. Rev. 40, 569 (1932).

[9] P. R. Ryason, Proposed Direct Test of the Uncertainty Principle, Phys. Rev. 115, 784 (1959).

[10] Marvin Chester and Roger Maynard, Microscopic Origin for the Apparent Uncertainty Principle Governing the Anomalous Attenuation of Third Sound, Phys. Rev. Lett. 29, 628 (1972).

[11] S. J. Putterman, R. Finkelstein, and I. Rudnick, Macroscopic Quantum Uncertainty Principle and Superfluid Hydrodynamics, Phys. Rev. Lett. 27, 1697 (1971).

[12] I. Kanter, Undecidability principle and the uncertainty principle even for classical systems, Phys. Rev. Lett. 64, 332 (1990) .

[13] David K. Hoffman and Donald J. Kouri, Hierarchy of Local Minimum Solutions of Heisenberg's Uncertainty Principle, Phys. Rev. Lett. 85, 5263 (2000).

[14] Robert Brandenberger and Pei-Ming Ho, Noncommutative spacetime, stringy spacetime uncertainty principle, and density fluctuations, Phys. Rev. D 66, 023517 (2002) .

[15] A. Bina, S. Jalalzadeh, and A. Moslehi, Quantum black hole in the generalized uncertainty principle framework, Phys. Rev. D 81, 023528 (2010) .

[16] C. Huang and Yong-Chang Huang, unification theory of classical statistical uncertainty relation and quantum uncertainty relation and its applications, Physics Letters A 375 (2011) 271–275.

[17] Pierre Nataf, Mehmet Dogan, and Karyn Le Hur, Heisenberg uncertainty principle as a probe of entanglement entropy: Application to superradiant quantum phase transitions, Phys. Rev. A 86, 043807 (2012).

[18] Giuseppe Vallone, Davide G. Marangon, Marco Tomasin, and Paolo Villoresi, Quantum randomness certified by the uncertainty principle, Phys. Rev. A 90, 052327 (2014).

[19] Smail Bougouffa and Zbigniew Ficek, Evidence of indistinguishability and entanglement determined by the energy-time uncertainty principle in a system of two strongly coupled bosonic modes, Phys. Rev. A 93, 063848 (2016).

[20] L. Perivolaropoulos, Cosmological horizons, uncertainty principle, and maximum length quantum mechanics, Phys. Rev. D 95, 103523 (2017).

[21] Shmuel Friedland, Vlad Gheorghiu, and Gilad Gour, Universal Uncertainty Relations, Phys. Rev. Lett. 111, 230401 (2013) .

[22] Andre C. Barato and Udo Seifert, Thermodynamic Uncertainty Relation for Biomolecular Processes, Phys. Rev. Lett. 114, 158101 (2015) .

[23] K. Yosida, Functional Analysis, Springer-Verlag, Berlin, 1980.

[24] G. H. Hardy, J. E. Littlewood and G. Polya, Inequalities, the second edition, Cambridge University Press, 1952.

[25] J. J. Sakurai, Jim Napolitano, Modern quantum mechanics, 2nd Edition, Addison-Wesley, 2011.

[26] Foster Ethan Kamer, Path Integral Approach to Quantum Physics: An Introduction, Springer, 1996; Khandekar D C et al, Path Integral Methods and Their Applications, WORLD SCIENTIFIC PUBLISHING (1993).

[27] Y. Deng, C. Huang, Yong-Chang Huang, Constraining the existence of magnetic monopoles by Dirac-dual electric charge renormalization effect under the Planck scale limit,

INTERNATIONAL JOURNAL OF MODERN PHYSICS, A31 (2016) 1650133.

[28] Lennard-Jones, J. E., On the Determination of Molecular Fields, Proc. R. Soc. Lond. A, 106 (738): (1924) 463–477; Barron, T. H. K.; Domb, C., On the Cubic and Hexagonal Close-Packed Lattices, Proceedings of the Royal Society of London. Series A, Mathematical and Physical Sciences, 227 (1171): (1955) 447–465,

[29] Peter Atkins and Julio de Paula, Atkins' Physical Chemistry (8th edn, W. H. Freeman), p. 637; D. C. Rapaport. The Art of Molecular Dynamics Simulation, Cambridge University Press, (2004).

[30] Reichl, L. E., A modern course in statistical physics, 3rd Edition, Weinheim: Wiley-VCH, 2009.